\documentclass[namedreferences]{SSRv}
\usepackage{epsfig}
\usepackage{graphicx}
\usepackage{times}
\usepackage{makeidx}

\def\dd{{\rm d}}
\def\etal{{\em et al.}}
\def\mat{{\rm mat}}
\def\de{{\rm de}}

\begin{document}

\begin{article}
\begin{opening}

\title{Fundamental constants and tests of general relativity -
       Theoretical and cosmological considerations}

\author{Jean-Philippe Uzan}

 \runningauthor{Jean-Philippe Uzan}
 \runningtitle{Fundamental constants and tests of general relativity}

\institute{Institut d'Astrophysique de Paris,
              UMR-7095 du CNRS, Universit\'e Pierre et Marie
              Curie,
              98 bis bd Arago, 75014 Paris (France)
              \email{uzan@iap.fr}}

\date{Received ; accepted }

\begin{abstract}
 The tests of the constancy of the fundamental constants are
 tests of the local position invariance and thus of the equivalence
 principle. We summarize the various constraints that have been
 obtained and then describe the connection between varying
 constants and extensions of general relativity. To finish, we
 discuss the link with cosmology, and more particularly with the
 acceleration of the Universe. We take the opportunity to
 summarize various possibilities to test general relativity (but
 also the Copernican principle) on cosmological scales.
\end{abstract}

\end{opening}
 \vspace{-0,5cm}
{\small

\section{Introduction}\label{sec_intro}

Physical theories usually introduce constants, i.e. numbers that
are not, and by construction can not be, determined by the theory
in which they appear. They are contingent to the theory and can
only be experimentally determined and measured.

These numbers have to be assumed constant for two reasons. First,
from a theoretical point of view, we have no evolution equation
for them (since otherwise they would be fields) and they cannot be
expressed in terms of other more fundamental quantities. Second,
from an experimental point of view, in the regimes in which the
theories in which they appear have been validated, they should be
constant at the accuracy of the experiments, to ensure the
reproducibility of experiments. This means that testing for the
constancy of these parameters is a test of the theories in which
they appear and allow to extend the knowledge of their domain of
validity.

Indeed, when introducing new, more unified or more fundamental,
theories the number of constants may change so that the list of
what we call fundamental constants is a time-dependent concept and
reflects both our knowledge and ignorance (Weinberg, 1983). Today,
gravitation is described by general relativity, and the three other
interactions and whole fundamental fields are described by the
standard model of particle physics. In such a framework, one has
22 unknown constants [the Newton constant, 6 Yukawa couplings for
the quarks and 3 for the leptons, the mass and vacuum expectation
value of the Higgs field, 4 parameters for the
Cabibbo-Kobayashi-Maskawa matrix, 3 coupling constants, a UV
cut-off to which one must add the speed of light and the Planck
constant; see e.g. Hogan (2000)].

Since any physical measurement reduces to the comparison of two
physical systems, one of them often used to realize a system of
units, it only gives access to dimensionless numbers. This implies
that only the variation of dimensionless combinations of the
fundamental constants can be measured and would actually also
correspond to a modification of the physical laws [see e.g. Uzan
(2003), Ellis and Uzan (2005)]. Changing the value of some
constants while letting all dimensionless numbers unchanged would
correspond to a change of units. It follows that from the 22
constants of our reference model, we can pick 3 of them to define
a system of units (such as e.g. $c$, $G$ and $h$ to define the
Planck units) so that we are left with 19 unexplained
dimensionless parameters, characterizing the mass hierarchy, the
relative magnitude of the various interactions etc.

Indeed, this number can change with time. For instance, we know
today that neutrinos have to be somewhat massive. This implies
that the standard model of particle physics has to be extended and
that it will involve at least 7 more parameters (3 Yukawa
couplings and 4 CKM parameters). On the other hand, this number
can decrease, e.g. if the non-gravitational interactions are
unified. In such a case, the coupling constants may be related to
a unique coupling constant $\alpha_U$ and a mass scale of
unification $M_U$ through
$$
 \alpha_i^{-1}(E) = \alpha_U^{-1} +
 \frac{b_i}{2\pi}\ln\frac{M_U}{E},
$$
where the $b_i$ are numbers which depends on the explicit model of
unification. This would also imply that the variations, if any, of
various constants will be correlated.

The tests of the constancy of fundamental constants take all their
importance in the realm of the tests of the equivalence principle
(Will, 1993). This principle, which states the universality of
free fall, the local position invariance and the local Lorentz
invariance, is at the basis of all metric theories of gravity and
implies that all matter fields are universally coupled to a unique
metric $g_{\mu\nu}$ which we shall call the physical metric,
$$
 S_{\rm matter}(\psi,g_{\mu\nu}).
$$
The dynamics of the gravitational sector is dictated by the
Einstein-Hilbert action
$$
 S_{\rm grav} = \frac{c^3}{16\pi G}\int\sqrt{-g_*}R_*\dd^4x.
$$
General relativity assumes that both metrics coincide,
$g_{\mu\nu}=g^*_{\mu\nu}$.

The test of the constancy of constants is a test of the local
position invariance hypothesis and thus of the equivalence
principle. Let us also emphasize that it is deeply related to the
universality of free fall since if any constant $c_i$ is a
space-time dependent quantity so will the mass of any test
particle so that it will experience an anomalous acceleration
$$
 \delta\vec a=\frac{\partial\ln m}{\partial c_i}\dot c_i\vec v,
$$
which is composition dependent [see Uzan (2003) for a review].

In particular, this allows to extend tests of the equivalence, and
thus tests of general relativity, on astrophysical scales. Such
tests are central in cosmology in which the existence of a dark
sector (dark energy and dark matter) is required to explain the
observations.

Necessity of theoretical physics in our understanding of
fundamental constants and on deriving bounds on their variation
is, at least, threefold:
\begin{enumerate}
 \item it is necessary to understand and to model the physical
 systems used to set the constraints. In particular one needs to
 determine the effective parameters that can be
 observationally constrained to a set of fundamental constants;
 \item it is necessary to relate and compare different constraints
 that are obtained at different space-time positions. This often
 requires a space-time dynamics and thus to specify a model;
 \item it is necessary to relate the variation of different
 fundamental constants through e.g. unification.
\end{enumerate}

This text summarizes these three aspects by first focusing, in
section~\ref{sec1}, on the various physical systems that have been
used, in section~\ref{sec2}, on the theories describing varying
constants (focusing on unification and on the link with the
universality of free fall). Section~\ref{sec3} summarizes the
links with cosmology where our current understanding of the
dynamics of the cosmic expansion calls for the introduction of
another constant: the cosmological constant, if not of a new
sector of physics, the ``dark energy''.

\section{Physical systems and constraints}\label{sec1}

\subsection{Physical systems}

The various physical systems that have been considered can be
classified in many ways.

First, we can classify them according to their look-back time and
more precisely their space-time position relative to our actual
position. This is summarized on Fig.~\ref{fig1} which represents
our past-light cone, the location of the various systems (in terms
of their redshift $z$) and the typical level at which they
constrain the time variation of the fine structure constant. This
systems include atomic clocks comparisons ($z=0$), the Oklo
phenomenon ($z\sim0.14$), meteorite dating ($z\sim0.43$), both
having a space-time position along the world line of our system
and not on our past-light cone, quasar absorption spectra
($z=0.2-4$), cosmic microwave background (CMB) anisotropy
($z\sim10^3$) and primordial nucleosynthesis (BBN, $z\sim10^8$).
Indeed higher redshift systems offer the possibility to set
constraints on an larger time scale, but at the prize of usually
involving other parameters such as the cosmological parameters.
This is particularly the case of the cosmic microwave background
and primordial nucleosynthesis, the interpretation of which
require a cosmological model [see Uzan (2003) for a review and
Uzan and Leclercq (2005) for a non-technical introduction].

The systems can also be classified in terms of the physics that
they involve in order to be interpreted (see Table~\ref{tab-sum}).
For instance, atomic clocks, quasar absorption spectra and the
cosmic microwave background require only to use quantum
electrodynamics to draw the primary constraints, so that these
constraints will only involve the fine structure constant
$\alpha$, the ratio between the proton-to-electron mass ratio
$\mu$ and the various gyromagnetic factors $g_I$. On the other
hand, the Oklo phenomenon, meteorite dating and nucleosynthesis
require nuclear physics and quantum chromodynamics to be
interpreted (see below).

\begin{table}[t]
\begin{center}
\begin{tabular}{p{2.8 cm}ccc}
 \hline\hline
 System  & Observable & Primary constraints  & Other hypothesis \\
 \hline
 Atomic clock      & $\delta\ln\nu$           &  $g_i,\alpha,\mu$ & - \\
 Oklo phenomenon   & isotopic ratio           &  $E_r$          & geophysical model \\
 Meteorite dating  & isotopic ratio           &  $\lambda$      &  -\\
 Quasar spectra    & atomic spectra           &  $g_p,\mu,\alpha$ & cloud properties\\
 21 cm             & $T_b$                    &  $g_p,\mu,\alpha$ & cosmological model\\
 CMB               & $T$                      &  $\mu,\alpha$     & cosmological model \\
 BBN               & light element abundances &  $Q_{np},\tau,m_e,m_N,\alpha,B_D$
          & cosmological model     \\
 \hline\hline
\end{tabular}
\caption{Summary of the systems considered to set constraints on
the variation of the fundamental constants. We summarize the
observable quantities (see text for details), the primary
constants used to interpret the data and the other hypotheses
required for this interpretation. [$\alpha$:fine structure
constant; $\mu$: electron-to-proton mass ratio; $g_i$:
gyromagnetic factor; $E_r$: resonance energy of the samarium-149;
$\lambda$: lifetime; $B_D$: deuterium binding energy; $Q_{np}$:
neutron-proton mass difference; $\tau$: neutron lifetime; $m_e$:
mass of the electron; $m_N$: mass of the nucleon].}
\label{tab-sum}
\end{center}
\end{table}

\begin{figure}%
 \centerline{\includegraphics[width=9cm]{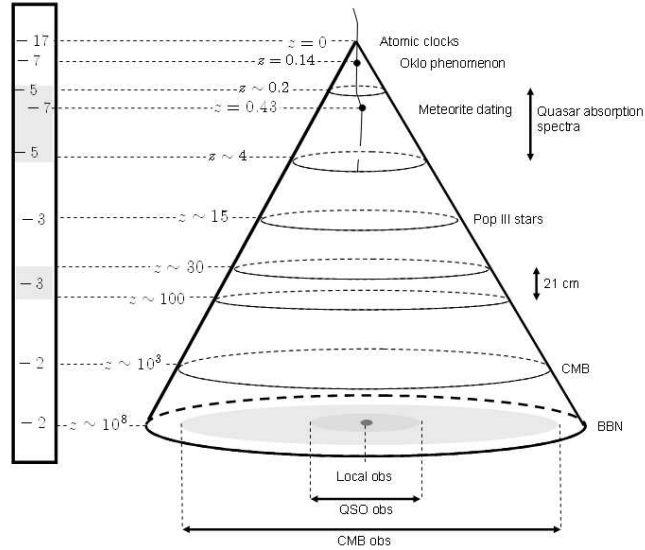}}
 \caption{Summary of the systems that have been used to probe the constancy of
  the fundamental constants and their position in a space-time diagram in which the cone
  represents our past light cone. The shaded areas represent the comoving space probed
  by different tests with respect to the largest scales probed by
  primordial nucleosynthesis.}
  \label{fig1}
\end{figure}

\subsection{Setting constraints}

For any system, setting constraints goes through several steps
that we sketch here.

First, any system allows us to derive an observational or
experimental constraint on an observable quantity $O(G_k,X)$ which
depends on a set of primary physical parameters $G_k$ and a set of
external parameters $X$, that usually are physical parameters that
need to be measured or constrained (e.g. temperature,...). These
external parameters are related to our knowledge of the physical
system and the lack of their knowledge is usually referred to as
systematic uncertainty.

From a physical model of the system, one can deduce the
sensitivities of the observables to an independent variation of
the primary physical parameters
\begin{equation}
 \kappa_{G_k} = \frac{\partial\ln O}{\partial\ln G_k}.
\end{equation}
As an example, the ratio between various atomic transitions can be
computed from quantum electrodynamics to deduce that the ratio of
two hyperfine-structure transition depends only on $g_I$ and
$\alpha$ while the comparison of fine-structure and
hyperfine-structure transitions depend on $g_I$, $\alpha$ and
$\mu$. For instance (Dzuba \etal~(1999); Karshenboim(2005))
$$
 \frac{\nu_{\rm Cs}}{\nu_{\rm Rb}} \propto \frac{g_{\rm Cs}}{g_{\rm
 Rb}}\alpha^{0.49},\qquad
 \frac{\nu_{\rm Cs}}{\nu_{\rm H}} \propto g_{\rm
 Cs}\mu\alpha^{2.83}.
$$

The primary parameters are usually not fundamental constants (e.g.
the resonance energy of the samarium $E_r$ for the Oklo
phenomenon, the deuterium binding energy $B_D$ for nucleosynthesis
etc.) The second step is thus to relate the primary parameters to
(a choice of) fundamental constants $c_i$. This would give a
series of relations (see e.g. M\"uller \etal~(2004))
\begin{equation}
 \Delta\ln G_k =\sum_i d_{ki}\Delta\ln c_i.
\end{equation}
The determination of the parameters $d_{ki}$ requires first to
choose the set of constants $c_i$ (do we stop at the masses of the
proton and neutron, or do we try to determine the dependencies on
the quark masses, or on the Yukawa couplings and Higgs vacuum
expectation value, etc.; see e.g. Dent \etal (2008) for various
choices) and also requires to deal with nuclear physics and the
intricate structure of QCD. In particular, the energy scales of
QCD, $\Lambda_{\rm QCD}$, is so dominant that at lowest order all
parameters scales as $\Lambda_{\rm QCD}^n$ so that the variation
of the strong interaction would not affect dimensionless
parameters and one has to take the effect of the quark masses.

As an example, the Oklo phenomenon allows to draw a constraint on
the value of the energy of the resonance. The observable $O$ is a
set of isotopic ratios that allow to reconstruct the average
cross-sections for the nuclear network that involves the various
isotopes of the samarium and gadolinium (this involves assumptions
about the geometry of the reactor, its temperature that falls into
$X$). It was argued (Damour and Dyson, 1996) on the basis of a
model of the samarium nuclei that the energy of the resonance is
mainly sensitive to $\alpha$ so that the only relevant parameter
is $ d_\alpha\sim -1.1\times10^7$. The level of the constraint
$-0.9\times10^{-7}<\Delta\alpha/\alpha<1.2\times10^{-7}$ that is
inferred from the observation is thus related to the sensitivity
$d_\alpha$.

\subsection{Constraints on the fine structure constant}

An extended discussion of the different constraints on the time
variation of the fine structure constant can be found in Uzan
(2003, 2004). We just summarize the state of the art in
Fig.~\ref{fig2} which depicts the constraints on
$\Delta\alpha/\alpha$ for different redshift bands. In the cases
where the constraints involve other constants, we have assumed for
simplicity that only $\alpha$ was allowed to vary. This is the
case in particular for quasar absorption spectra (see the
contribution by P. Petitjean in this volume).

The typical constraints on cosmological timescales (of order of
10~Gyr) is $\Delta\alpha/\alpha<10^{-6}$ which would correspond to
a constraint $\Delta\alpha/\alpha<10^{-16}\,{\rm yr}^{-1}$ if one
assumes (but there a priori no reason to support it) that the rate
of change is constant over time (i.e. $\alpha(t)$ is a linear
function of the cosmic time). For comparison, under the same
hypothesis, Oklo would give the constraint
$\Delta\alpha/\alpha<5\times 10^{-17}\,{\rm yr}^{-1}$. But indeed
these constraints are complementary since they concern different
spacetime positions.

\begin{figure}%
 \centerline{\includegraphics[width=11cm]{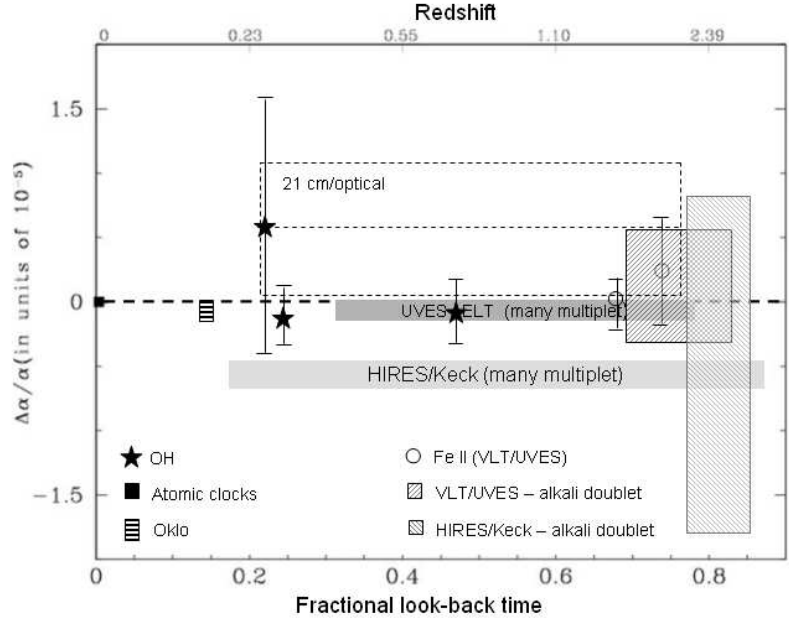}}
 \caption{Constraints on the variation of the fine structure constant obtained
 for the various physical systems as a function of redshift or look-back
 time, assuming the standard $\Lambda$CDM model.}\label{fig2}
\end{figure}

\subsection{Unification and correlated variations}

In the context of the unification of the fundamental interactions,
it is expected that the variations of the various constants are
not independent. By understanding these correlations we can set
stronger constraints at the expense of being more model-dependent.
We only illustrate this on the example of BBN, along the lines of
Coc \etal~(2007).

The BBN theory predicts the production of the light elements in
the early universe: the abundances synthesized rely on the balance
between the expansion of the universe and the weak interaction
rates which control the neutron to proton ratio at the onset of
BBN (see Peter and Uzan (2005) for a textbook introduction).
Basically, the abundance of helium-4 is well approximated by
\begin{equation}
 Y_p=2\frac{(n/p)_{\rm f}\exp(-t_{\rm N}/\tau)}{1+(n/p)_{\rm f}\exp(-t_{\rm N}/\tau)}
\end{equation}
where $(n/p)_{\rm f}=\exp{(-Q_{np}/kT_{\rm f})}$ is the neutron to
proton ratio at the freeze-out time determined (roughly) by
$G_F^2(kT_{\rm f})^5=\sqrt{GN}(kT_{\rm f})^2$, $N$ being the
number of relativistic degrees of freedom; $Q_{np}=m_{\rm
n}-m_{\rm p}$, $\tau$ is the neutron lifetime, $G_F$ the Fermi
constant and $t_{\rm N}$ the time after which the photon density
becomes low enough for the photo-dissociation of the deuterium to
be negligible. As a conclusion, the predictions of BBN involve a
large number of fundamental constants. In particular, $t_{\rm N}$
depends on the deuterium binding energy and on the
photon-to-baryon ratio, $\eta$. Besides, one needs to include the
effect of the fine structure constant in the Coulomb barriers
(Bergstr\"om \etal, 1999). For a different analysis of the effect of
varying fundamental constants on BBN predictions see e.g. M\"uller
\etal~(2004), Landau \etal~(2006), Coc \etal~(2007), Dent
\etal~(2007).

It follows that the predictions of the BBN are mainly dependent of
the effective parameters $G_k=
(G,\alpha,m_e,\tau,Q_{np},B_D,\sigma_i)$ while the external
parameters are mainly the cosmological parameters
$X=(\eta,h,N_\nu,\Omega_i)$. The dependence of the predictions on
the independent variation of each of these parameters can be
determined and it was found (Flambaum and Shuryak (2002), Coc
\etal~(2007): Fig.~3) that the most sensitive parameter is the
deuterium binding energy, $B_D$ (see also Dent \etal~(2007) for a
similar analysis including more parameters). The helium and
deuterium data allow to set the constraints
$$
 -7.5\times10^{-2}<\frac{\Delta B_D}{B_D}< 6.5\times10^{-2},\quad
 -8.2\times10^{-2}<\frac{\tau}{\tau}< 6\times10^{-2},
$$
$$
 -4\times10^{-2}<\frac{\Delta Q_{np}}{Q_{np}}< 2.7\times10^{-2}
$$
on the independent variations of these parameter. More
interestingly (and more speculative!) a variation of $B_D$ in the
range $(-7.5,-4)\times10^{-2}$ would be compatible with the helium
and deuterium constraint while reconciling the spectroscopically
determined lithium-7 abundance (Coc \etal, 2007) with its expected
value from WMAP.

In a second step, the parameters $G_k$ can be related to a smaller
set of fundamental constants, namely the fine structure constant
$\alpha$, the Higgs VEV $v$, the Yukawa couplings $h_i$ and the
QCD scale $\Lambda_{\rm QCD}$ since $Q_{np}=m_n-m_p = \alpha
a\Lambda_{\rm QCD} + (h_d-h_u)v$, $m_e=h_ev$, $\tau_n =
G_F^2m_e^5f(Q/m_e)$ and $G_F=1/\sqrt{2}v$. The deuterium binding
energy can be expressed in terms of $h_s$, $v$ and $\Lambda_{\rm
QCD}$ (Flambaum and Shuryak, 2003) using a sigma nuclear model or
in terms of the pion mass (Epelbaum \etal, 2003). Assuming that
all Yukawa couplings vary similarly, the set of parameters $G_k$
reduces to the set of constants $\{\alpha,v,h,\Lambda_{\rm QCD}\}$
(again in units of the Planck mass).

Several relations between these constants can however be found.
For instance, in grand-unified models the low-energy expression of
$\Lambda_{\rm QCD}$,
$$
 \Lambda_{\rm QCD} = \mu\left(\frac{m_cm_bm_t}{\mu^3}\right)^{2/27}
 \exp\left[-\frac{2\pi}{9\alpha_3(\mu)}\right]
$$
for $\mu>m_t$ yields a relation between $\{\alpha,v,h,\Lambda_{\rm
QCD}\}$ so that one actually has only 3 independent constants.
Then, in all models in which the weak scale is determined by
dimensional transmutation, changes in the Yukawa coupling $h_t$
will trigger changes in $v$ (Ibanez and Ross, 1982). In such
cases, the Higgs VEV can be written as
$$
 v = M_p\exp\left[-\frac{8\pi^2c}{h_t^2}\right],
$$
where $c$ is a constant of order unity. I follows that we are left
with only 2 independent constants. This number can even be reduced
to 1 in the case where one assumes that the variation of the
constants is trigger by an evolving dilaton (Damour and Polyakov
(1994); Campbell and Olive (1995)). At each stage, one reduces the
number of constants, and thus the level of the constraints, at the
expense of some model dependence. In the latter case, it was shown
that BBN can set constraints of the order of
$|\Delta\alpha/\alpha|<4\times10^{-5}$ (Coc \etal, 2007).

\section{Theories with ``varying constants''}\label{sec2}

\subsection{Making a constant dynamical}

The question of whether the constants of nature may be dynamical
goes back to Dirac (1937) who expressed, in his ``Large Number
hypothesis", the opinion that very large (or small) dimensionless
universal constants cannot be pure mathematical numbers and must
not occur in the basic laws of physics. In particular, he stressed
that the ratio between the gravitational and electromagnetic
forces between a proton and an electron, $Gm_{\rm e}m_{\rm
p}/e^2\sim10^{-40}$ is of the same order as the inverse of the age
of the universe in atomic units, $e^2H_0/m_{\rm e}c^3$. He stated
that these were not pure numerical coincidences but instead that
these big numbers were not pure constants but reflected the state
of our universe. This led him to postulate that $G$
varies\footnote{Dirac hypothesis can also be achieved by assuming
that $e$ varies as $t^{1/2}$. Indeed this reflects a choice of
units, either atomic or Planck units. There is however a
difference: assuming that only $G$ varies violates the strong
equivalence principle while assuming a varying $e$ results in a
theory violating the Einstein equivalence principle. It does not
mean we are detecting the variation of a dimensionful constant but
simply that either $e^2/\hbar c$ or $Gm_{\rm e}^2/\hbar c$ is
varying.} as the inverse of the cosmic time. Diracs' hypothesis is
indeed not a theory and it was shown later that a varying constant
can be included in a Lagrangian formulation as a new dynamical
degree of freedom so that one gets both a new dynamical equation
of evolution for this degree of freedom and a modification of the
other field equations with respect to their form derived under the
hypothesis it was constant.

Let us illustrate this on the case of scalar-tensor theories, in
which gravity is mediated not only by a massless spin-2 graviton
but also by a spin-0 scalar field that couples universally to
matter fields (this ensures the universality of free fall). In the
Jordan frame, the action of the theory takes the form
\begin{eqnarray}\label{actionJF}
  S &=&\int \frac{\dd^4 x }{16\pi G_*}\sqrt{-g}
     \left[F(\varphi)R-g^{\mu\nu}Z(\varphi)\varphi_{,\mu}\varphi_{,\nu}
        - 2U(\varphi)\right]+ S_{\rm matter}[\psi;g_{\mu\nu}]
\end{eqnarray}
where $G_*$ is the bare gravitational constant. This action
involves three arbitrary functions ($F$, $Z$ and $U$) but only two
are physical since there is still the possibility to redefine the
scalar field. $F$ needs to be positive to ensure that the graviton
carries positive energy. $S_{\rm matter}$ is the action of the
matter fields that are coupled minimally to the metric
$g_{\mu\nu}$. In the Jordan frame, the matter is universally
coupled to the metric so that the length and time as measured by
laboratory apparatus are defined in this frame.

It is useful to define an Einstein frame action through a
conformal transformation of the metric $g_{\mu\nu}^* =
F(\varphi)g_{\mu\nu}$. In the following all quantities labelled by
a star (*) will refer to Einstein frame. Defining the field
$\varphi_*$ and the two functions $A(\varphi_*)$ and
$V(\varphi_*)$ (see e.g. Esposito-Far\`ese and Polarski, 2001) by
$$
 \left(\frac{\dd\varphi_*}{\dd\varphi}\right)^2
              = \frac{3}{4}\left(\frac{\dd\ln F(\varphi)}{\dd\varphi}\right)^2
                  +\frac{1}{2F(\varphi)},
                  \quad
 A(\varphi_*) = F^{-1/2}(\varphi),\quad
 2V(\varphi_*)= U(\varphi) F^{-2}(\varphi),
$$
the action (\ref{actionJF}) reads as
\begin{eqnarray}
 S &=& \frac{1}{16\pi G_*}\int \dd^4x\sqrt{-g_*}\left[ R_*
        -2g_*^{\mu\nu} \partial_\mu\varphi_*\partial_\nu\varphi_*
        - 4V\right]+ S_{\rm matter}[A^2g^*_{\mu\nu};\psi].
\end{eqnarray}
The kinetic terms have been diagonalized so that the spin-2 and
spin-0 degrees of freedom of the theory are perturbations of
$g^*_{\mu\nu}$ and $\varphi_*$ respectively.

The action~(\ref{actionJF}) defines an effective gravitational
constant $G_{\rm eff} = G_*/F = G_*A^2$. This constant does not
correspond to the gravitational constant effectively measured in a
Cavendish experiment. The Newton constant measured in this
experiment is $G_{\rm cav} = G_*A_0^2(1+\alpha_0^2)$ where the
first term, $G_*A_0^2$ corresponds to the exchange of a graviton
while the second term $G_*A_0^2\alpha_0^2$ is related to the long
range scalar force. The gravitational constant depends on the
scalar field and is thus dynamical.

The post-Newtonian parameters can be expressed in terms of the
values of $\alpha$ and $\beta$ today as
\begin{equation}
 \gamma^{\rm PPN} - 1 = -\frac{2\alpha_0^2}{1+\alpha^2_0},\qquad
 \beta^{\rm PPN} - 1 =\frac{1}{2}
 \frac{\beta_0\alpha_0^2}{(1+\alpha_0^2)^2}.
\end{equation}
The Solar system constraints imply $\alpha_0$ to be very small,
typically $\alpha_0^2<10^{-5}$ while $\beta_0$ can still be large.
Binary pulsar observations~(Esposito-Far\`ese, 2005) impose that
$\beta_0>-4.5$ and that $\dot G/G<10^{-12}\,{\rm yr}^{-1}$.

\subsection{General dangers}

Given the previous discussion, it seems a priori simple to cook up
a theory that will describe a varying fine structure constant by
coupling a scalar field to the electromagnetic Faraday tensor as
\begin{equation}
 S = \int\left[\frac{R}{16\pi G} - 2(\partial_\mu\phi)^2
 -\frac{1}{4}B(\phi)F_{\mu\nu}^2 \right]\sqrt{-g}\dd^4 x
\end{equation}
so that the fine structure will evolve according to
$\alpha=B^{-1}$.

Such an simple implementation may however have dramatic
implications. In particular, the contribution of the
electromagnetic binding energy to the mass of any nucleus can be
estimated by the Bethe-Weiz\"acker formula so that
$$
 m_{(A,Z)}(\phi) \supset 98.25\,\alpha(\phi)\,\frac{Z(Z-1)}{A^{1/3}}\,\hbox{MeV}.
$$
This implies that the sensitivity of the mass to a variation of
the scalar field is expected to be of the order of
\begin{equation}
 f_{(A,Z)} = \partial_\phi m_{(A,Z)}(\phi) \sim 10^{-2}
 \frac{Z(Z-1)}{A^{4/3}} \alpha'(\phi).
\end{equation}
It follows that the level of the violation of the universality of
free fall is expected to be of the level of $
\eta_{12}\sim10^{-9}X(A_1,Z_1;A_2,Z_2)(\partial_\phi\ln B)^2_0$.
Since the factor $X(A_1,Z_1;A_2,Z_2)$ typically ranges as
$\mathcal{O}(0.1-10)$, we deduce that $(\partial_\phi\ln B)_0$ has
to be very small for the Solar system constraints to be satisfied.
It follows that today the scalar field has to be very close to the
minimum of the coupling function $\ln B$.

Let us mention that such coupling terms naturally appear when
compactifying a higher-dimensional theory. As an example, let us
recall the compactification of a 5-dimensional Einstein-Hilbert
action (Peter and Uzan (2005), chapter~13)
$$
 S=\frac{1}{12\pi^2 G_5}\int\bar R\sqrt{-\bar g}\dd^5 x.
$$
Decomposing the 5-dimensional metric $\bar g_{AB}$ as
$$
 \bar g_{AB} = \left(
\begin{array}{cc}
  g_{\mu\nu}+\frac{A_\mu A_\nu}{M^2}\phi^2 & \frac{A_\mu}{M}\phi^2\\
  \frac{A_\nu}{M}\phi^2 & \phi^2 \\
\end{array}
 \right),
$$
where $M$ is a mass scale, we obtain
\begin{equation}
 S=\frac{1}{16\pi G}\int\left(R - \frac{\phi^2}{4M^2}F^2\right)\phi\sqrt{-g}\dd^4
 x,
\end{equation}
where the 4-dimensional gravitational constant is $G=3\pi
G_5/4\int\dd y$. The scalar field couples explicitly to the
kinetic term of the vector field and cannot be eliminated by a
redefinition of the metric: this is the well-known conformal
invariance of electromagnetism in four dimensions. Such a term
induces a variation of the fine structure constant as well as a
violation of the universality of free-fall. Such dependencies of
the masses and couplings are generic for higher-dimensional
theories and in particular string theory. It is actually one of
the definitive predictions for string theory that there exists a
dilaton, that couples directly to matter (Taylor and Veneziano,
1988) and whose vacuum expectation value determines the string
coupling constants (Witten, 1984).

For example, in type I superstring theory, the 10-dimensional dilaton
couples differently to the gravitational and Yang-Mills sectors
because the graviton is an excitation of closed strings while the
Yang-Mills fields are excitations of open strings. For small
values of the volume of the extra-dimensions, a T-duality makes
the theory equivalent to a 10-dimensional theory with Yang-Mills
fields localized on a D3-brane. When compactified on an orbifold,
the gauge fields couple to fields $M_i$ living only at these
orbifold points with couplings $c_i$ which are not universal.
Typically, one gets that $M_4^2=\hbox{e}^{-2\Phi}V_6M_I^8$ while
$g_{YM}^{-2}= \hbox{e}^{-2\Phi}V_6M_I^6 +c_iM_i$. Unfortunately,
the 4-dimensional effective couplings depend on the version of the
string theory, on the compactification scheme and on the dilaton.

\subsection{Ways out}

While the tree-level predictions of string theory seem to be in
contradiction with experimental constraints, many mechanisms can
reconcile it with experiment. In particular, it has been claimed
that quantum loop corrections to the tree-level action may modify
the coupling function in such a way that it has a minimum~(Damour
and Polyakov, 1994). As explained in the former paragraph, the
dilaton needs to be close to the minimum of the coupling function
in order for the theory to be compatible with the universality of
free fall. In the case of scalar-tensor theories, it was shown
that when the coupling function enjoys such a minimum, the theory
is naturally attracted toward general relativity (Damour and
Nordtvedt, 1993). The same mechanism will apply if all the
coupling functions have the same minimum (see the contribution by
T. Damour in this volume for more details). In that
particular model the mass of any nuclei will typically be of the
form
$$
 m_i(\phi) = \Lambda_{QCD}(\phi)\times\left(1+
 a^q\frac{m_q}{\Lambda_{QCD}} + a^e\alpha\right),
$$
where $a^q$ and are $a^e$ are sensitivities. It follows that
composition independent effects (i.e. $|\gamma^{\rm PPN}-1|$,
$|\beta^{\rm PPN}-1|$, $\dot G/G$) and composition dependent
effects ($\eta$, $\dot\alpha$, $\dot\mu$) will be of the same
order of magnitude, dictated by the difference of the value of the
dilaton today compared to its value at the minimum of the coupling
function.

Another possibility is to invoke an environmental dependence, as
can be implemented in scalar-tensor theories by the chameleon
mechanism~(Khoury and Weltman, 2004) which invokes a potential
with a minimum that does not coincide with the one of the coupling
function.

\section{Links with cosmology}\label{sec3}

The comparison of various constraints requires a full cosmological
model. In particular, one cannot assume that the time variation of
the constant is linear either with time or redshift (as used to
compare the typical magnitudes of the constraints in \S~2.3).
Besides, cosmology tends to indicate that a new constant, the
cosmological constant, has to be included in our description. We
briefly address these two points by focusing on the tests that can
be performed on cosmological scales and on the status of our
cosmological model.

\subsection{Cosmological evolution}

The cosmological dynamics is central to apply the least coupling
principle. Since the dilaton is attracted towards the minimum of
the coupling function during its cosmological evolution,
deviations from general relativity are expected to be larger in
the early universe. In particular BBN can set bounds on the
deviation from general relativity that are stronger than those
obtained from Solar system experiments (Damour and Pichon (1999), Coc
\etal~(2006)).

\subsection{About dark energy}

The construction of any cosmological model relies on 4 main
hypotheses: (H1) a theory of gravity; (H2) a description of the
matter contained in the Universe and their non-gravitational
interactions; (H3) symmetry hypothesis; (H4) an hypothesis on the
global structure, i.e. the topology of the Universe.

These hypotheses are not on the same footing since H1 and H2 refer
to the physical theories. These two hypotheses are however not
sufficient to solve the field equations and we must make an
assumption on the symmetries (H3) of the solutions describing our
Universe on large scales while H4 is an assumption on some global
properties of these cosmological solutions, with same local
geometry.

Our reference cosmological model is the $\Lambda$CDM model. It
assumes that gravity is described by general relativity (H1), that
the Universe contains the fields of the standard model of particle
physics plus some dark matter and a cosmological constant, the
latter two having no physical explanation at the moment. It also
deeply involves the Copernican principle as a symmetry hypothesis
(H3), without which the Einstein equations usually can not been
solved, and usually assumes that the spatial sections are simply
connected (H4). H2 and H3 imply that the description of the standard
matter reduces to a mixture of a pressureless and a radiation
perfect fluids. This model is compatible with all astronomical
data which roughly indicates that $ \Omega_{\Lambda0} \simeq
0.73$, $\Omega_{\rm mat0} \simeq 0.27,$ and $\Omega_{K0} \simeq
0$. Cosmology thus roughly imposes that $|\Lambda_0|\leq H_0^2$,
that is $\ell_\Lambda \leq H_0^{-1} \sim 10^{26}\,\mathrm{m} \sim
10^{41}\,\mathrm{GeV}^{-1}$. Notice that it is disproportionately
large compared  to the natural scale fixed by the Planck length
\begin{equation}
 \rho_{\Lambda_0}<10^{-120}M_{\rm
 Pl}^4\sim10^{-47}\,\mathrm{GeV}^4\,
\end{equation}
at the heart of the cosmological constant problem.

The assumption that the Copernican principle holds, and the fact
that it is so central in drawing our conclusion on the
acceleration of the expansion, splits the investigation into two
avenues. Either we assume that the Copernican principle holds and
we have to modify the laws of fundamental physics or we abandon
the Copernican principle, hoping to explain dark energy without
any new physics but at the expense of living in a particular place
in the Universe. In the former case, the models can be classified
in terms of 4 universality classes summarized on
Fig.~\ref{fig-name3}).

\begin{figure}%
 \centerline{\includegraphics[width=9cm]{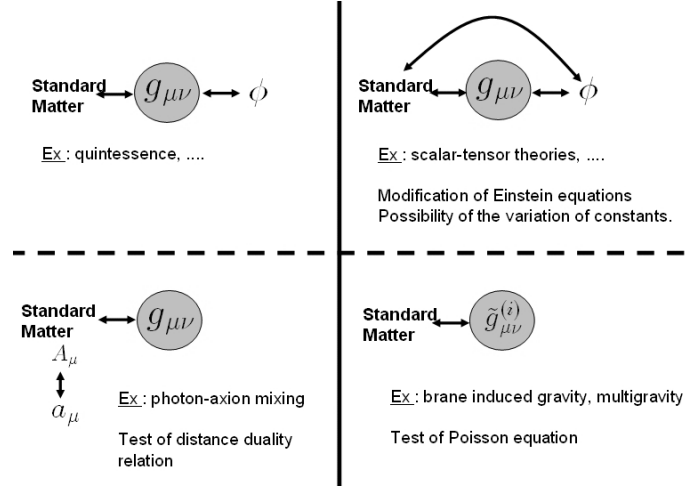}}
 \caption{Summary of the different classes of physical dark energy models. As discussed in the text,
 various tests can be designed to distinguish between them. The classes differ according
 to the nature of the new degrees of freedom and their couplings.
 Left column accounts for models
 where gravitation is described by general relativity while right column models
 describe a modification of general relativity. In the upper line classes, the new fields
 dominate the matter content of the universe at low redshift.
 Upper-left models (class A) consist of
 models in which a new kind of gravitating matter is introduced. In the upper-right
 models (class C), a light field induces a long-range force so that gravity is
 not described by a massless spin-2 graviton only. In this class, Einstein equations
 are modified and there may be a variation of the fundamental
 constants. The lower-right models (class D) correspond to models in which there
 may exist an infinite number of new degrees of freedom, such as in some class of braneworld
 scenarios. These models predict a modification of the Poisson
 equation on large scales. In the last class (lower-left, class B), the
 distance duality relation may be violated. From Uzan (2007).}\label{fig-name3}
\end{figure}

The goal in this section is to summarize some attempts to test
both the Copernican principle and general relativity on
cosmological scales.

\subsection{Beyond the Copernican principle}

The Copernican principle implies that the spacetime metric reduces
to a single function, the scale factor $a(t)$ that can be Taylor
expanded as $a(t)=a_0+H_0(t-t_0) - \frac{1}{2} q_0 H_0^2 (t-t_0)^2
+\ldots$. It follows that the conclusions that the cosmic
expansion is accelerating ($q_0<0$) does not involve any
hypothesis about the theory of gravity (other than the one that
the spacetime geometry can be described by a metric) or the matter
content, as long as this principle holds.

While isotropy around us seems well established observationally,
homogeneity is more difficult to test. The possibility, that we
may be living close to the center of a large under-dense region
has sparked considerable interest, because such models can
successfully match the magnitude-redshift relation of type Ia
supernovae without dark energy (see the contribution by S. Sarkar
in this volume).

The main difficulty in testing the Copernican principle lies in
the fact that most observations are located on our past
light-cone. Recently, it was realized that cosmological
observations may however provide such a test (Uzan \etal, 2008).
It exploits the time drift of the redshift that occurs in any
expanding spacetime, as first pointed out in the particular case
of Robertson-Walker spacetimes for which it takes the form $\dot z
= (1+z)H_0 - H(z)$. Such an observation would give informations on
the dynamics outside the past light-cone since it compares the
redshift of a given source at two times and thus on two infinitely
close past light-cones (see Fig.~\ref{fig4}-right). It follows
that it contains informations about the spacetime structure
along the worldlines of the observed sources that must be
compatible with the one derived from the data along the past
light-cone.

For instance, in a spherically symmetric spacetime, the expression
depends on the shear, $\sigma(z)$, of the congruence of the
wordlines of the comoving observers evaluated along our past
light-cone (Uzan \etal, 2008),
$$
 \dot z = (1+z)H_0 - H(z) -\frac{1}{\sqrt{3}}\sigma(z)\ .
$$
It follows that, when combined with other distance data, it allows
to determine the shear on our past light-cone and one can check
whether it is compatible with zero, as expected for any
Robertson-Walker spacetime.

In a RW universe, one can go further and determine a consistency
relation between several observables since $H^{-1}(z)= D'(z)\left[
1 + \Omega_{K0}H_0^2D^2(z) \right]^{-1/2}$, where a prime stands
for $\partial_z$ and $D(z)=D_L(z)/(1+z)$; this relation is
independent of the Friedmann equations. It follows that in any
Robertson-Walker spacetime the {\it consistency relation},
$$
 1 + \Omega_{K0}H_0^2\left(\frac{D_L(z)}{1+z}\right)^2 -
 \left[H_0(1+z) -\dot z(z)
 \right]^2\left[\frac{\dd}{\dd
 z}\left(\frac{D_L(z)}{1+z}\right)\right]^2=0,
$$
between observables must hold whatever the matter content and the
field equations, since it derives from pure kinematical relations
that do not rely on the dynamics.

$\dot z(z)$ has a typical amplitude of order $\delta z\sim
-5\times10^{-10}$ on a time scale of $\delta t= 10$~yr, for  a
source at redshift $z=4$. This measurement is challenging, and
impossible with present-day facilities. However, it was recently
revisited in the context of Extremely Large Telescopes (ELT),
arguing they could measure velocity shifts of order $\delta v\sim
1-10\ {\rm cm/s}$ over a 10 years period from the observation of
the Lyman-$\alpha$ forest. It is one of the science drivers in
design of the CODEX spectrograph~(Pasquini \etal, 2005) for the
future European ELT. Indeed, many effects, such as proper motion
of the sources, local gravitational potential, or acceleration of
the Sun may contribute to the time drift of the redshift. It was
shown~(Uzan \etal, 2008), however, that these contributions can be
brought to a 0.1\% level so that the cosmological redshift is
actually measured.

Let us also stress that another idea was also proposed
recently~(Goodman, 1995; Caldwell and Stebbins, 2008; Clarkson
\etal~2008, see the contribution by R. Caldwell in this
volume). This idea is based on the distortion of the Planck spectrum
of the cosmic microwave background.

\begin{figure}%
 \centerline{\includegraphics[width=8cm]{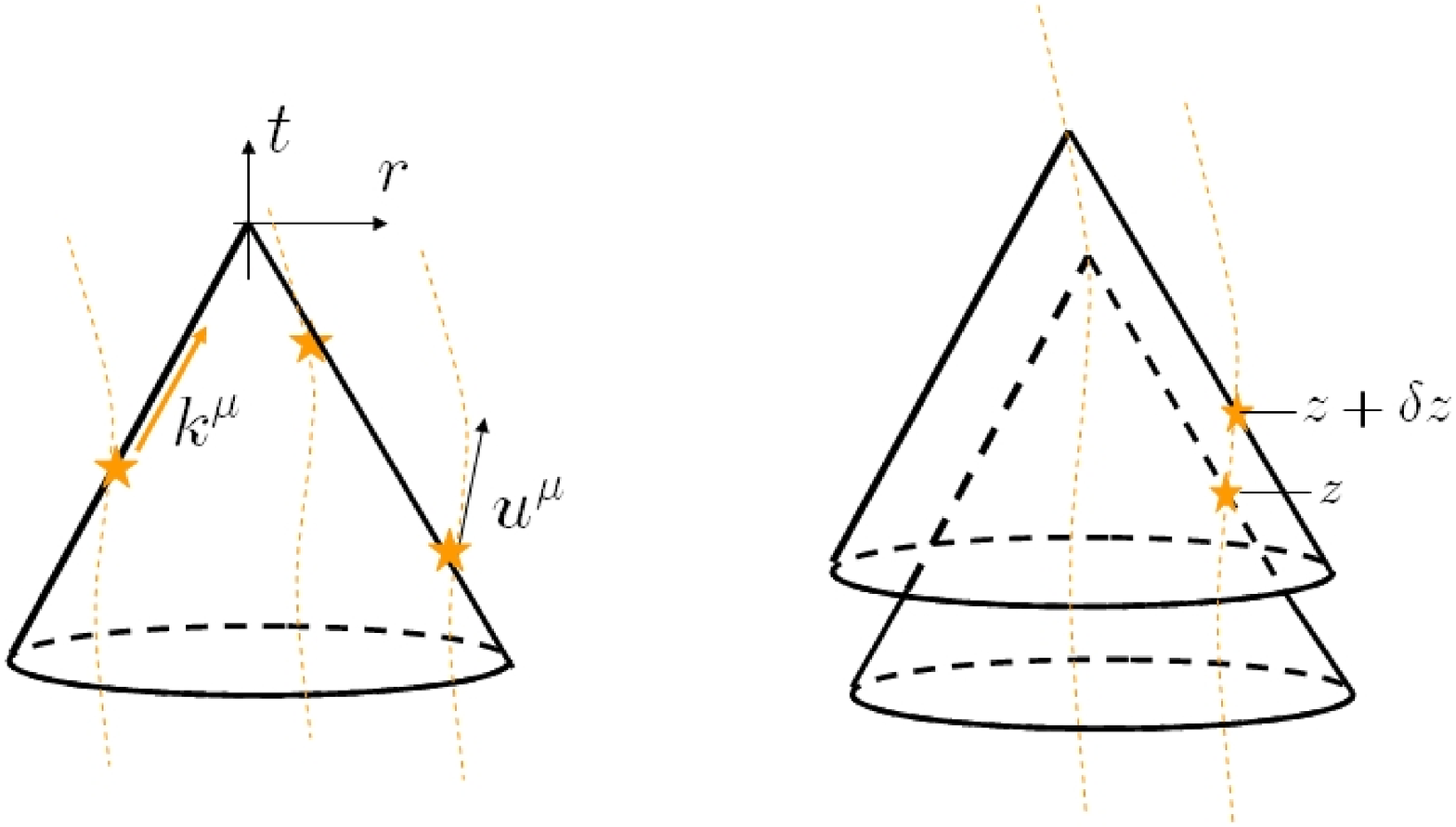}}
 \caption{{\it Left}: Most low-redshift data are localized on our past light-cone.
    In a non-homogeneous spacetime there is no direct relation between
    the redshift that is observed and the cosmic time, needed to reconstruct
    the expansion history.
    {\it Right}: The time drift of the redshift allows to extract information
    about two infinitely close past light-cones. $\delta z$ depends
    on the proper motions of the observer and the sources as well
    as the spacetime geometry.}
    \label{fig4}
\end{figure}

\subsection{Testing General Relativity on astrophysical scales}

Extracting constraints on deviations from GR on cosmological
scales is difficult because large scale structures entangle the
properties of matter and gravity. On sub-Hubble scales, one can,
however, construct tests reproducing those in the Solar system.
For instance, light deflection is a test of GR because we can
measure independently the deflection angle and the mass of the
Sun.

On sub-Hubble scales, relevant for the study of the large-scale
structure, the Einstein equations reduce to the Poisson equation
\begin{equation}\label{Peq}
\Delta\Psi = 4\pi G\rho_\mat a^2
\delta_\mat=\frac{3}{2}\Omega_\mat H^2 a^2\delta_\mat,
\end{equation}
relating the gravitational potential and the matter density
contrast.

As first pointed out by Uzan and Bernardeau (2001), this relation
can be tested on astrophysical scales, since the gravitational
potential and the matter density perturbation can be measured
independently from the use of cosmic shear measurements and galaxy
catalogs. The test was recently implemented with the CFHTLS-weak
lensing data and the SDSS data to conclude that the Poisson
equation holds observationally to about 10~Mpc~(Dor\'e \etal,
2007). The main limitation in the applicability of this test is
due to the biasing mechanisms (i.e. the fact that galaxies do not
necessarily trace faithfully the matter field) even if it is
thought to have no significant scale dependence at such scales.

\subsubsection{Toward a post-$\Lambda$CDM formalism}

The former test of the Poisson equation exploits one rigidity of
the field equations on sub-Hubble scales. It can be improved by
considering the full set of equations.

Assuming that the metric of spacetime takes the form
\begin{equation}
 \dd s^2 = -(1+2\Phi)\dd t^2 + (1-2\Psi)a^2\gamma_{ij}\dd x^i\dd x^j
\end{equation}
on sub-Hubble scales, the equation of evolution reduces to the
continuity equation  $\delta_\mat' + \theta_\mat =0$, where
$\theta$ is the divergence of the velocity perturbation and a
prime denotes a derivative with respect to the conformal time, the
Euler equation $\theta_\mat' +{\cal H}\theta_\mat= -\Delta\Phi$,
where ${\cal H}$ is the comoving Hubble parameter,the Poisson
equation~(\ref{Peq}) and $\Phi=\Psi$.

These equations imply many relations between the cosmological
observables. For instance, decomposing $\delta_\mat$ as
$D(t)\epsilon(x)$ where $\epsilon$ encodes the initial conditions,
the growth rate $D(t)$ evolves as $ \ddot D + 2H\dot D -4\pi
G\rho_\mat D = 0$. This equation can be rewritten in terms of
$p=\ln a$ as time variable~(Peter and Uzan, 2005) and considered
not as a second order equation for $D(t)$ but as a first order
equation for $H^2(a)$
$$
 (H^2)'+ 2\left(\frac{3}{a} + \frac{D''}{D'} \right)H^2 =
 3\frac{\Omega_{\mat0}H_0^2 D}{a^2 D'}
$$
where a prime denotes a derivative with respect to $p$. It can be
integrated as~(Chiba and Nakamura, 2007)
\begin{equation}\label{ddd}
 \frac{H^2(z)}{H_0^2} = 3\Omega_{\mat0}\left(\frac{1+z}{D'(z)}\right)^2
 \int\frac{D}{1+z}(-D')\dd z.
\end{equation}
This exhibits a rigidity between the growth function and the
Hubble parameter. In particular the Hubble parameter determined
from background data and from perturbation data using
Eq.~(\ref{ddd}) must agree. This was used in the analysis of Wang
\etal (2007).

Another relation exists between $\theta_\mat$ and $\delta_\mat$.
The Euler equation implies that
$\theta_\mat=-\beta(\Omega_{\mat0},\Omega_{\Lambda0})\delta_\mat$,
with $\beta(\Omega_{\mat0},\Omega_{\Lambda0})\equiv {\dd\ln
D(a)}/{\dd\ln a}$.

We conclude that the perturbation variables are not independent
and the relation between them are inherited from some assumptions
on the dark energy. Phenomenologically, we can generalize the
sub-Hubble equations to
\begin{eqnarray}
 \delta_\mat' + \theta_\mat =0\qquad\qquad\qquad\quad,
 &&\theta_\mat' +{\cal H}\theta_\mat= -\Delta\Phi + S_\de,\\
  -k^2\Phi = 4\pi G F(k,H) \delta_\mat + \Delta_\de,
 &&\Delta(\Phi-\Psi) = \pi_\de.
\end{eqnarray}
We assume that there is no production of baryonic matter so that
the continuity equation is left unchanged. $S_\de$ describes the
interaction between dark energy and standard matter. $\Delta_\de$
characterizes the clustering of dark energy, $F$ accounts for a
scale dependence of the gravitational interaction and $\pi_\de$ is
an effective anisotropic stress. It is clear that the $\Lambda$CDM
corresponds to $(F,\pi_\de,\Delta_\de,S_\de) = (1,0,0,0)$. The
expression of $(F,\pi_\de,\Delta_\de,S_\de)$ for quintessence,
scalar-tensor, $f(R)$ and DGP models and more generally for models
of the classes A-D can be found in Uzan (2007).

From an observational point of view, weak lensing survey gives
access to $\Phi+\Psi$, galaxy maps allow to reconstruct
$\delta_g=b\delta_\mat$ where $b$ is the bias, velocity fields
give access to $\theta$. In a $\Lambda$CDM, the correlations
between these observables are not independent since, for instance
$\langle\delta_g\delta_g\rangle = b^2\langle\delta_\mat^2\rangle$,
$\langle\delta_g\theta_m\rangle = -b\beta
\langle\delta_\mat^2\rangle$ and $\langle\delta_g\kappa\rangle =
8\pi G\rho_\mat a^2 b\langle\delta_\mat^2\rangle$. Various ways of
combining these observables have been proposed, construction of
efficient estimators and forecast for possible future space
missions designed to make these tests as well as the possible
limitations (arising e.g. from non-linear bias, the effect of
massive neutrinos or the dependence on the initial conditions) are
now being extensively studied~(Zhang \etal, 2007; Jain and Zhang,
2007; Amendola \etal, 2008; Song and Koyama, 2008).

\section{Conclusion}

The study of fundamental constants provides tests of general
relativity that can be extended on astrophysical scales. These
constrains can be useful in our understanding of the origin of the
acceleration of the cosmic expansion. They can be combined with
tests of general relativity based on the large scale structure of
the universe.

In the coming future, we can hope to obtain new constraints from
population~III stars ($z\sim15$) and 21cm absorption
($z\sim30-100$) and improved constraints from QSO by about a
factor 10 (and probably more with ELT).

Interpreting the coupled variations of various constants is also
challenging since it usually implies to deal with nuclear physics
and QCD. It also offers a window on unification mechanisms and on
string theory.

All these aspects make the study of fundamental constants a lively
and promising topic.

}
\end{article}

\end{document}